\documentclass[a4paper]{jpconf}

\usepackage{graphicx}

\begin{document}

\title{Four questions for quantum-classical hybrid theory~\footnote{Main parts   
of author's talk at  
{\it ``Heinz von Foerster Congress / Emergent Quantum Mechanics''} (Vienna, Nov. 2011) 
are in Ref.\,\cite{me11}; questions are considered here which were raised 
afterwards or in related correspondence.}}  

\author{Hans-Thomas Elze}

\address{Dipartimento di Fisica ``Enrico Fermi'',  
        Largo Pontecorvo 3, I-56127 Pisa, Italia }

\ead{elze@df.unipi.it}

\begin{abstract} 
Four questions are discussed which may be 
addressed to any proposal of a quantum-classical hybrid theory  
which concerns the direct coupling of classical and quantum mechanical degrees of freedom. 
In particular, we consider the formulation of hybrid dynamics presented recently 
in Ref.\,\cite{me11}. This linear theory 
differs from the nonlinear ensemble theory of Hall and Reginatto, but shares with it 
to fulfil all consistency requirements discussed in the literature, while earlier 
attempts failed. -- Here, we additionally ask: Does the theory allow for  
superposition, separable, and entangled states originating in the quantum mechanical sector? 
Does it allow for ``Free Will'', as introduced, in this context, 
by Di\'osi \cite{Diosi11}? Is it local? Does it provide hints for the emergence 
of quantum mechanics from dynamics beneath?  
\end{abstract}

\section{Introduction}
The hypothetical direct coupling of quantum mechanical and classical degrees of 
freedom -- {\it ``hybrid dynamics''} -- departs from quantum mechanics.  
It has been researched at lenght for practical as well as theoretical reasons. 
In particular, the standard Copenhagen interpretation entails the unresolved 
measurement problem which, together with the fact that quantum mechanics needs  
interpretation, in order to be operationally well defined, may indicate that 
it deserves amendments. It has been recognized early on that 
a theory which {\it dynamically} bridges the quantum-classical divide should have an 
impact on the measurement problem \cite{Sudarshan123} as well as on  
attempts to describe consistently the interaction between quantum matter and classical 
spacetime \cite{BoucherTraschen}.  
   
Numerous works have appeared, in order to formulate hybrid dynamics in a satisfactory 
way. Generally, they showed deficiencies in one or another respect. 
Which has led to various no-go theorems in view of a list of desirable properties 
or consistency requirements, see, for example, 
Refs.\,\cite{CaroSalcedo99,DiosiGisinStrunz}:  
\begin{itemize} 
\item Conservation of energy. 
\item Conservation and positivity of probability. 
\item Separability of quantum and classical subsystems in the absence of their interaction, 
recovering the correct quantum and classical equations of motion. 
\item Consistent definitions of states and observables; existence of a Lie bracket structure 
on the algebra of observables that suitably generalizes  
Poisson and commutator brackets. 
\item Existence of canonical transformations generated by the observables; 
invariance of the classical sector under canonical transformations 
performed on the quantum sector only and {\it vice versa}. 
\item Existence of generalized Ehrenfest relations ({\it i.e.} the 
correspondence limit) which, 
for bilinearly coupled classical and quantum oscillators,  
are to assume the form of the classical equations of motion  
(``Peres-Terno benchmark'' test \cite{PeresTerno}). 
\end{itemize} 

These issues have been reviewed in recent works by Hall and Reginatto where  
they have introduced the first theory of hybrid dynamics that 
conforms with the points listed above \cite{HallReginatto05,Hall08,ReginattoHall08}. 
Their ensemble theory is based on configuration space, which requires a certain 
nonlinearity of the action functional from which it is derived and entails 
effects that might allow to falsify this proposal experimentally. 

We have proposed an alternative theory 
of hybrid dynamics based on notions of phase space \cite{me11}. 
This is partly motivated by work on related topics of general linear dynamics and 
classical path integrals \cite{EGV11,EGV10} and extends work by Heslot, who  
demonstrated that quantum mechanics can entirely be rephrased in the language and 
formalism of classical analytical mechanics \cite{Heslot85}. 
Introducing unified notions 
of states on phase space, observables, canonical transformations, and a generalized 
quantum-classical Poisson bracket, this has led to an intrinsically 
linear hybrid theory, which fulfils the above consistency requirements. 

Objects that somehow reside between classical and quantum mechanics 
have been described recently also in a statistical theory, based on very different premises 
than the hybrid theories considered here \cite{Zwitters}. It would be interesting to 
uncover any relation, if it exists. 

In any case, it must be emphasized that it is also of great practical importance to 
better understand quantum-classical hybrids appearing in quantum mechanical approximation 
schemes. These typically address many-body systems or interacting fields, 
which are naturally separable into quantum and classical subsystems, for example, representing fast and slow degrees of freedom 
(keywords: Born-Oppenheimer approximation, mesoscopic systems, ``semiclassical 
quantum gravity''); for references see Ref.\,\cite{me11}.  
 
Furthermore, concerning the hypothetical emergence of quantum mechanics from some  
coarse-grained deterministic dynamics (see, for example, Refs.\,\cite{tHooft10,Elze09a,Adler} 
with numerous references to earlier work), the quantum-classical backreaction problem seems  
to appear in a new form, namely regarding the interplay of fluctuations among underlying deterministic and emergent quantum mechanical degrees of freedom. Which can be 
rephrased succinctly as the question: {\it ``Can quantum mechanics be seeded?''}

Besides constructing the quantum-classical hybrid formalism and showing how 
it conforms with the above consistency requirements, we discussed  
the possibility to have classical-environment induced decoherence, quantum-classical 
backreaction, a deviation from the Hall-Reginatto proposal in presence 
of translation symmetry, and closure of the algebra of hybrid observables \cite{me11}.

Presently, we add four questions to accompany the above consistency 
requirements: 
\begin{itemize} 
\item Does the hybrid theory allow for the superposition principle, for separable and for 
entangled states originating in the quantum mechanical sector? 
\item Does it allow for ``Free Will'', a notion introduced, in this context, 
by Di\'osi \cite{Diosi11}? 
\item Does it describe a local interaction of classical and quantum mechanical subsystems 
which evolve in spacetime? 
\item Does it provide hints concerning the study of emergence 
of quantum mechanics from deterministic dynamics beneath? 
\end{itemize}
which may be addressed to any model intended to describe quantum-classical hybrids.  

The remainder of the paper is organized as follows.
In Section~2., we collect some of the results of Ref.\,\cite{me11}, which will 
be useful in the following. In Sections~3. to 6., we discuss, if not answer, 
the questions just listed with respect to the hybrid theory 
proposed in Ref.\,\cite{me11}, followed by concluding remarks in Section~7. 

\section{Linear quantum-classical hybrid dynamics -- a summary} 
In the following two subsections, we will present some important  
results drawn from the exposition of classical Hamiltonian 
mechanics and its generalization incorporating quantum mechanics 
by Heslot \cite{Heslot85}. 
This will form the starting point of our brief review of the hypothetical 
direct coupling between quantum and classical degrees of freedom, as 
developed in Ref.\,\cite{me11}; the reader familiar with the earlier 
derivations may directly pass to Section~3. 

\subsection{Classical mechanics} 
The evolution of a {\it classical} object is described with respect to its $2n$-dimensional 
phase space, which is identified as its {\it state space}. A real-valued regular 
function on the state space defines an {\it observable}, {\it i.e.}, a differentiable function 
on this smooth manifold. 

Darboux's theorem shows that there always exist (local) systems of so-called 
{\it canonical coordinates}, commonly denoted by $(x_k,p_k),\; k=1,\dots ,n$, 
such that the {\it Poisson bracket} of any pair of observables $f,g$ assumes 
the standard form \cite{Arnold}: 
\begin{equation}\label{PoissonBracket} 
\{ f,g \}\; =\; 
\sum_k\Big (\frac{\partial f}{\partial_{x_k}}\frac{\partial g}{\partial_{p_k}}
-\frac{\partial f}{\partial_{p_k}}\frac{\partial g}{\partial_{x_k}}\Big ) 
\;\;. \end{equation} 
This is consistent with $\{ x_k,p_l\}=\delta_{kl}$, $\{ x_k,x_l\} =\{ p_k,p_l\}=0,\; 
k,l=1,\dots ,n$, and reflects the bilinearity, antisymmetry, derivation-like 
product formula, and Jacobi identity which define a Lie bracket operation, 
$f,g\rightarrow\{ f,g\}$, mapping a pair of observables to an observable.  

Compatibility with the Poisson bracket structure restricts general transformations 
${\cal G}$ of the state space to so-called {\it canonical transformations} 
which do not change the physical properties of the object under study; e.g., a translation, 
a rotation, a change of inertial frame, or evolution in time. Such ${\cal G}$ 
induces a change of an observable, $f\rightarrow {\cal G}(f)$, and is an automorphism 
of the state space compatible with its Poisson bracket structure.
Due to the Lie group structure of the set of canonical transformations, it is sufficient 
to consider infinitesimal transformations generated by the elements of the 
corresponding Lie algebra. Then, an {\it infinitesimal transformation} ${\cal G}$ is 
{\it canonical}, if and only if for any observable $f$ the map $f\rightarrow {\cal G}(f)$ 
is given by $f\rightarrow f'=f+\{ f,g\}\delta\alpha$, with some observable $g$, 
the so-called {\it generator} of ${\cal G}$, and $\delta\alpha$ an infinitesimal real 
parameter. 

Thus, for the canonical coordinates, in particular, an infinitesimal canonical transformation 
amounts to: 
\begin{eqnarray}\label{xcan} 
x_k&\rightarrow &x_k'=x_k+\frac{\partial g}{\partial p_k}\delta\alpha 
\;\;, \\ [1ex] \label{pcan}  
p_k&\rightarrow &p_k'=p_k-\frac{\partial g}{\partial x_k}\delta\alpha 
\;\;, \end{eqnarray} 
employing the Poisson bracket given in Eq.\,(\ref{PoissonBracket}). 

This analysis shows the fundamental relation between observables and generators 
of infinitesimal canonical transformations in classical Hamiltonian mechanics. 

\subsection{Quantum mechanics} 
An important achievement of Heslot's work is the realization that the previous analysis 
can be generalized and applied to quantum mechanics; this concerns 
the dynamical aspects as well as the notions of states, canonical transformations, and 
observables. 

To begin with, we recall that the Schr\"odinger equation and its adjoint can be 
derived from a well known action principle \cite{me11}. Which shows immediately 
that they have the character of Hamiltonian equations of motion.  

To this must be added 
the {\it normalization condition} for the state vector $|\Psi\rangle$: 
\begin{equation}\label{normalization} 
{\cal C}:=\langle\Psi (t)|\Psi (t)\rangle\stackrel{!}{=}\mbox{constant}\equiv 1 
\;\;, \end{equation} 
which is an essential ingredient of the probability interpretation associated with 
state vectors. -- Adding here that state vectors that differ by an {\it unphysical constant 
phase} are to be identified, we are reminded that the {\it quantum mechanical state space} 
is formed by the rays of an underlying Hilbert space, {\it i.e.}, forming a complex 
projective space. 

\subsubsection{The oscillator representation}
Quantum mechanical evolution can be described by a unitary transformation, 
$|\Psi (t)\rangle =\hat U(t-t_0)|\Psi (t_0)\rangle$, with $U(t-t_0)=\exp [-i\hat H(t-t_0)]$, 
which formally solves the Schr\"odinger equation. 
It follows immediately that a stationary state, characterized by 
$\hat H|\phi_i\rangle =E_i|\phi_i\rangle$, with a real energy eigenvalue $E_i$,  
performs a simple harmonic motion, {\it i.e.}, 
$|\psi_i(t)\rangle =\exp [-iE_i(t-t_0)]|\psi_i(t_0)\rangle
\equiv\exp [-iE_i(t-t_0)]|\phi_i\rangle$. Henceforth, we assume a denumerable set of 
eigenstates of the Hamilton operator.

With the Hamiltonian character of the underlying 
equation(s) of motion, the harmonic motion suggests 
to introduce what we called {\it oscillator representation}. 
We consider 
the expansion of any state vector with respect to a complete 
orthonormal basis, $\{ |\Phi_i\rangle\}$:  
\begin{equation}\label{oscillexp} 
|\Psi\rangle =\sum_i|\Phi_i\rangle (X_i+iP_i)/\sqrt 2 
\;\;, \end{equation} 
where the generally time dependent expansion coefficients are written 
in terms of real and imaginary parts, $X_i,P_i$. Employing this expansion, allows to   
evaluate what will serve as the {\it Hamiltonian function}, 
{\it i.e.}, ${\cal H}:=\langle\Psi |\hat H|\Psi\rangle$: 
\begin{equation}\label{HamiltonianQM1} 
{\cal H}=\frac{1}{2}\sum_{i,j}\langle\Phi_i|\hat H|\Phi_j\rangle (X_i-iP_i)(X_j+iP_j) 
=:{\cal H}(X_i,P_i) 
\;\;. \end{equation} 
Choosing especially the set of energy eigenstates, $\{ |\phi_i\rangle\}$, 
as basis, we obtain:     
\begin{equation}\label{HamiltonianQM2} 
{\cal H}(X_i,P_i)=\sum_i\frac{E_i}{2}(P_i^{\;2}+X_i^{\;2}) 
\;\;, \end{equation} 
hence the name {\it oscillator representation}.  
The reasoning leading to this result indicates that $(X_i,P_i)$ 
can play the role of {\it canonical coordinates} in the description of a 
quantum mechanical object and its evolution with respect to the state space. 

This interpretation is substantiated by the fact that 
the Schr\"odinger equation is recovered by 
evaluating $|\dot\Psi\rangle =\sum_i|\Phi_i\rangle (\dot X_i+i\dot P_i)/\sqrt 2$ 
according to Hamilton's equations of motion, using the Hamiltonian function of 
Eq.\,(\ref{HamiltonianQM1}) or (\ref{HamiltonianQM2}). --   
Furthermore, the {\it constraint} of Eq.\,(\ref{normalization}) becomes: 
\begin{equation}\label{oscillnormalization} 
{\cal C}(X_i,P_i)=\frac{1}{2}\sum_i(X_i^{\;2}+P_i^{\;2})\stackrel{!}{=}1 
\;\;. \end{equation} 
Thus, the vector with components given by the canonical coordinates 
$(X_i,P_i),\; i=1,\dots ,N$, is 
constrained to the surface of a $2N$-dimensional sphere with radius $\sqrt 2$. 
This constraint presents a major difference to classical Hamiltonian mechanics. 

Similarly as in Subsection~2.1., it is natural to introduce 
also here a {\it Poisson bracket} for any two {\it observables} on the 
{\it spherically compactified state space}, {\it i.e.} real-valued regular functions $F,G$ of 
the coordinates $(X_i,P_i)$:     
\begin{equation}\label{QMPoissonBracket} 
\{ F,G \}\; =\; 
\sum_i\Big (\frac{\partial F}{\partial_{X_i}}\frac{\partial G}{\partial_{P_i}}
-\frac{\partial F}{\partial_{P_i}}\frac{\partial G}{\partial_{X_i}}\Big ) 
\;\;. \end{equation} 
Then, the Hamiltonian acts as the generator 
of time evolution of any observable $O$: 
\begin{equation}\label{evolution}  
\frac{\mbox{d}O}{\mbox{d}t}=\partial_tO+\{ O,{\cal H} \} 
\;\;, \end{equation} 
and one verifies that the constraint, 
Eq.\,(\ref{oscillnormalization}), is conserved under the Hamiltonian flow:  
\begin{equation}\label{constraint} 
\frac{\mbox{d}{\cal C}}{\mbox{d}t}=\{ {\cal C},{\cal H} \}=0 
\;\;. \end{equation} 

It remains to demonstrate the {\it compatibility} of the notion of 
{\it observable} introduced here -- as in classical mechanics -- with the one adopted in 
quantum mechanics. This concerns, in particular, the implementation of {\it canonical 
transformations} and the role of observables as their generators.    

\subsubsection{Canonical transformations and quantum observables} 
The Hamiltonian function has been introduced as observable in the 
Eq.\,(\ref{HamiltonianQM1}) which provides a direct relation to the corresponding quantum  
observable, namely the expectation value of the self-adjoint Hamilton operator. 
This is indicative of the general structure to be discussed now.    

Refering to Refs.\,\cite{me11,Heslot85} for all details, we 
summarize here the main points:  
\\ \noindent $\bullet$ 
A) {\it Compatibility of unitary transformations and Poisson structure.} -- 
The canonical transformations discussed in Section~2.1. represent 
automorphisms of the classical state space which are compatible with the Poisson brackets. 
In quantum mechanics automorphisms of the Hilbert space are implemented by 
unitary transformations. This implies a transformation of the canonical 
coordinates here, {\it i.e.}, of the expansion coefficients $(X_i,P_i)$ introduced in 
Eq.\,(\ref{oscillexp}). Analyzing this, one finds that  
the fundamental Poisson brackets remain invariant under unitary transformations. 
Thus, {\it unitary transformations on Hilbert space are canonical transformations 
on the $(X,P)$ state space}.   
\\ \noindent $\bullet$ 
B) {\it Self-adjoint operators as observables.} -- 
Any infinitesimal unitary transformation $\hat U$ can be generated by a self-adjoint operator 
$\hat G$, such that: 
\begin{equation}\label{Uinfini} 
\hat U=1-i\hat G\delta\alpha 
\;\;, \end{equation} 
which will lead to the quantum mechanical relation between an observable and 
a self-adjoint operator. In fact, straightforward calculation shows that in the present case 
we have: 
\begin{eqnarray}\label{Xcan} 
X_i&\rightarrow &X_i'=X_i+\frac{\partial \langle\Psi |\hat G|\Psi\rangle}
{\partial P_i}\delta\alpha 
\;\;, \\ [1ex] \label{Pcan}  
P_i&\rightarrow &P_i'=P_i-\frac{\partial \langle\Psi |\hat G|\Psi\rangle}
{\partial X_i}\delta\alpha 
\;\;. \end{eqnarray} 
Then, the  
relation between an observable $G$, defined in analogy to Section~2.1., and a self-adjoint 
operator $\hat G$ can be inferred from Eqs.\,(\ref{Xcan})--(\ref{Pcan}) within a few 
steps: 
\begin{equation}\label{goperator} 
G(X_i,P_i)=\langle\Psi |\hat G|\Psi\rangle 
\;\;, \end{equation} 
by comparison with the classical result.  
We find that a {\it real-valued regular function $G$ of the state is an observable, if 
and only if there exists a self-adjoint operator $\hat G$ such that Eq.\,(\ref{goperator}) 
holds}. This implies that {\it all quantum observables are quadratic forms} 
in the $X_i$'s and $P_i$'s, which are essentially fewer than in the corresponding classical 
case; see, however, Subsection~3.2 for the generalization necessitated by interacting 
quantum-classical hybrids. 
\\ \noindent $\bullet$ 
C) {\it Commutators as Poisson brackets.} -- 
Relation (\ref{goperator}) between observables and self-adjoint operators together 
with the Poisson bracket (\ref{QMPoissonBracket}) allow to 
demonstrate the important result: 
\begin{equation}\label{QMPBComm} 
\{ F,G\}=\langle\Psi |\frac{1}{i}[\hat F,\hat G]|\Psi\rangle 
\;\;, \end{equation}  
with both sides of the equality considered as functions of the variables $X_i,P_i$ 
and with the commutator defined as usual. This shows that 
the {\it commutator is a Poisson bracket with respect to the $(X,P)$ state space} and 
relates the algebra of observables, in the sense of the classical construction of Section~2.1.,  
to the algebra of self-adjoint operators in quantum mechanics.   
\\ \noindent $\bullet$ 
D) {\it Normalization, phase arbitrariness, and admissible observables.} -- 
Coming back to the normalization condition $\langle\Psi |\Psi\rangle\stackrel{!}{=}1$, 
which compactifies the state space, cf. the constraint Eq.\,(\ref{oscillnormalization}), 
it must be preserved under infinitesimal canonical transformations, since it belongs 
to the structural characteristics of the state space. This leads to a constraint on 
admissible observables, which turns out to be compatible with the requirement that 
observables $G$ are invariant under an infinitesimal 
phase transformation $|\Psi\rangle\;\rightarrow\; |\Psi\rangle\cdot\exp (i\delta\theta )$, 
with constant $\delta\theta$. Conversely, assuming this {\it phase invariance} 
of observables, we 
recover that Hilbert space vectors differing by a constant phase are 
indistinguishable and represent the same physical state. 

We note that any observable $G$ with an expansion as in Eq.\,(\ref{HamiltonianQM1})  
automatically satisfies the invariance requirements of item D).  
Explicit calculation shows:  
\begin{equation}\label{constraintinvar1}
\{ {\cal C},G\}=\sum_j\Big (
\frac{\partial G}{\partial P_j}X_j-\frac{\partial G}{\partial X_j}P_j\Big )=0
\;\;, \end{equation} 
assuming that: 
\begin{equation}\label{Gexp} 
G(P_i,X_i):=\langle\Psi |\hat G|\Psi\rangle 
=\frac{1}{2}\sum_{i,j}G_{ij}(X_i-iP_i)(X_j+iP_j)
\;\;, \end{equation} 
and where $G_{ij}:=\langle\Phi_i|\hat G|\Phi_j\rangle =G_{ji}^\ast$, for a 
self-adjoint operator $\hat G$. 

In conclusion, quantum mechanics shares with classical mechanics an even dimensional state 
space, a Poisson structure, and a related algebra of observables. Yet it  
differs essentially by a restricted set of observables and the requirements 
of phase invariance and normalization, which compactify the underlying Hilbert space 
to the complex projective space formed by its rays.  

\subsection{Quantum-classical Poisson bracket and separability}
The result of aligning classical and quantum mechanics in the way summarized above 
suggests to introduce a {\it generalized Poisson bracket} for 
observables defined on the Cartesian product state space of CL (classical) {\it and} 
QM (quantum mechanical) sectors of a hybrid:  
\begin{eqnarray}\label{GenPoissonBracket} 
\{ A,B\}_\times &:=&\{ A,B\}_{\mbox{\scriptsize CL}}+\{ A,B\}_{\mbox{\scriptsize QM}}
\\ [1ex] \label{GenPoissonBracketdef} 
&:=&\sum_k\Big (\frac{\partial A}{\partial_{x_k}}\frac{\partial B}{\partial_{p_k}}
-\frac{\partial A}{\partial_{p_k}}\frac{\partial B}{\partial_{x_k}}\Big )+  
\sum_i\Big (\frac{\partial A}{\partial_{X_i}}\frac{\partial B}{\partial_{P_i}}
-\frac{\partial A}{\partial_{P_i}}\frac{\partial B}{\partial_{X_i}}\Big ) 
\;\;, \end{eqnarray} 
for any two observables $A,B$. It is bilinear and antisymmetric, leads to a derivation-like 
product formula and obeys the Jacobi identity. 

Let us say {\it an observable ``belongs'' to the CL (QM) sector, if it is  
constant with respect to the canonical coordinates of the QM (CL) sector}. -- 
Then, the generalized Poisson bracket has the additional important properties: 
\begin{itemize} 
\item It reduces to the Poisson brackets introduced 
in Eqs.\,(\ref{PoissonBracket}) and (\ref{QMPoissonBracket}), respectively,   
for pairs of observables that belong {\it either} to the CL {\it or} the QM sector. 
\item It reduces to the appropriate one of the former brackets, 
if one of the observables belongs only to either one of the two sectors. 
\item It reflects the {\it separability} of CL and QM sectors, 
since $\{ A,B\}_\times =0$, if $A$ and $B$ belong to different sectors. 
\end{itemize}  

The physical relevance of separability is this: 
{\it If a canonical tranformation 
is performed on the QM (CL) sector only, then all observables that belong to the 
CL (QM) sector remain invariant.}

\section{Superposition, separable and entangled states from the QM sector?} 
Whether our description of quantum-classical hybrids 
incorporates the possibility of superposition, 
separable, and entangled states originating in the QM sector  
can be answered right away in the affirmative. This follows from the 
fact that it represents quantum mechanics exactly, in the absence of 
a hybrid coupling, as we have shown \cite{me11}. This continues to hold, as long 
as the dynamical variables pertaining to the CL sector can be considered 
as external parameters for the QM sector. However, further comment is 
necessary when the QM-CL hybrid is 
fully dynamical and we will come to this in due course.   

\subsection{Definition and interpretation of hybrid states} 
It may be useful to look from a different angle at the question how different kinds of 
quantum states enter the description of hybrids. --  
We recall that the hybrid density 
$\rho$ has been introduced in Ref.\,\cite{me11} 
as the expectation with respect to a given state vector of a 
self-adjoint, positive semi-definite, trace normalized density operator    
$\hat\rho$ \cite{me11}:  
\begin{equation}\label{rhodens}
\rho (x_k,p_k;X_i,P_i):=\langle\Psi |\hat\rho (x_k,p_k)|\Psi\rangle
=\frac{1}{2}\sum_{i,j}\rho_{ij}(x_k,p_k)(X_i-iP_i)(X_j+iP_j)
\;\;, \end{equation} 
using the oscillator expansion, Eq.\,(\ref{oscillexp}), and   
$\rho_{ij}(x_k,p_k):=\langle\Phi_i|\hat\rho (x_k,p_k)|\Phi_j\rangle 
=\rho_{ji}^\ast (x_k,p_k)$. It describes a {\it quantum-classical hybrid ensemble} 
by a real-valued, positive semi-definite, normalized, and possibly time dependent 
regular function, the probability distribution $\rho$, on the Cartesian 
product state space canonically coordinated by $2(n+N)$-tuples $(x_k,p_k;X_i,P_i)$; 
the variables $x_k,p_k,\; k=1,\dots ,n$ and $X_i,P_i,\; i=1,\dots ,N$ are reserved 
for the CL and QM sector, respectively.

Furthermore, 
expanding $\hat\rho$ in terms of its eigenstates,   
$\hat\rho =\sum_jw_j|j\rangle\langle j|$, one obtains: 
\begin{eqnarray}\label{rhointerpr1}
\rho (x_k,p_k;X_i,P_i)&=&\sum_jw_j(x_k,p_k)\mbox{Tr}(|\Psi\rangle\langle\Psi |j\rangle\langle j|) 
\\ [1ex] \label{rhointerpr2}
&=&\sum_jw_j(x_k,p_k)|\langle j|\Psi\rangle |^2  
\;\;,  \end{eqnarray} 
with $0\leq w_j\leq 1$ and $\sum_j\int\Pi_l(\mbox{d}x_l\mbox{d}p_l)w_j(x_k,p_k)=1$. -- 
This suggests that $\rho (x_k,p_k;X_i,P_i)$, when properly  
normalized, is the {\it probability density to find in the hybrid ensemble the QM state} 
$|\Psi\rangle$, parametrized by $X_i,P_i$ through Eq.\,(\ref{oscillexp}), {\it together with  
the CL state} given by a point in phase space, specified by the coordinates $(x_k,p_k)$.   

Clearly, this interpretation does not depend on whether $\hat\rho$ stands for 
any {\it pure state}, which might be a coherent superposition of other pure states in  
the Hilbert space of the QM sector, or for a {\it mixed state}, i.e., a probabilistic 
superposition of pure states.  

Note, in particular, how a probabilistic mixture of two densities, 
$\rho :=p\rho^{(1)}+(1-p)\rho^{(2)}$, with $0\leq p\leq 1$, can be represented according to 
Eq.\,(\ref{rhointerpr2}): 
\begin{equation}\label{mixture} 
\rho (x_k,p_k;X_i,P_i)=\sum_j\Big (pw_j(x_k,p_k)^{(1)}+(1-p)w_j(x_k,p_k)^{(2)}\Big ) 
|\langle j|\Psi\rangle |^2  
\;\;,  \end{equation} 
which is consistent with the interpretation given. -- It must be emphasized that 
our definition of the density, Eq.\,(\ref{rhodens}), is not suitable to ask 
questions like {\it ``What is the probability to find in the hybrid ensemble a given 
QM mixed state?''}. It does not contain sufficient information for such, more 
general purposes. 
  
However, the observations made allow us to address the cases of 
{\it separable and entangled states} relevant for a composite quantum system as part 
of a hybrid. For simplicity, we treat a bi-partite QM sector; 
the generalization for multi-partite systems follows by induction. -- Generally,  
two {\it QM systems} A {\it and} B {\it are defined to be separable}, if their state can 
be prepared as a statistical mixture of tensor product states \cite{DiosiBook}: 
\begin{equation}\label{separable} 
\hat\rho^{(\mbox{AB})}=\sum_lw_l\hat\rho^{(\mbox{A})}_l\otimes\hat\rho^{(\mbox{B})}_l\;\;,\;\;\; 
\mbox{with}\;\;\sum_lw_l=1\;\;,\;\;\;w_l\geq 0
\;\;. \end{equation} 
If this is not the case, A and B are said to be in an {\it entangled state}.  
If A and B are separable, then there exist at most classical but no quantum correlations 
between them. For a pure state, this is equivalent to the statement that it is separable 
if and only if it is of the form $|\Psi^{(\mbox{AB})}\rangle =|\Psi^{(\mbox{A})}\rangle 
\otimes |\Psi^{(\mbox{B})}\rangle$; here, the sum in Eq.\,(\ref{separable}) has only 
one nonvanishing term. 

Considering {\it separable states}, generally, we can adapt 
Eqs.\,(\ref{rhointerpr1})--(\ref{rhointerpr2}) to the case at hand by expanding 
each one of the self-adjoint factors $\hat\rho^{(\mbox{A})}_l$ and $\hat\rho^{(\mbox{B})}_l$ in the 
definition, Eq.\,(\ref{separable}), in terms of its eigenstates, e.g., 
$\hat\rho^{(\mbox{A})}_l =
\sum_{j_l}w^{(\mbox{A})}_{j_l}|\mbox{A};j_l\rangle\langle\mbox{A};j_l|$.    
Inserting these expansions, the Eq.\,(\ref{rhointerpr2}) is replaced by: 
\begin{eqnarray}\label{rhointerpr2sep} 
\rho (x_k,p_k;X_i,P_i)&=&
\sum_{l,j_l,j'_l}w_l(x_k,p_k)w^{(\mbox{A})}_{j_l}(x_k,p_k)w^{(\mbox{B})}_{j'_l}(x_k,p_k)
|\langle j_l,j'_l|\Psi\rangle |^2  
\\ [1ex] \label{rhointerpr2sep1} 
&=&\sum_{j,j'}w^{(\mbox{A})}_{j}(x_k,p_k)w^{(\mbox{B})}_{j'}(x_k,p_k)
|\langle j,j'|\Psi\rangle |^2   
\;\;,  \end{eqnarray} 
where $|j_l,j'_l\rangle :=|A;j_l\rangle\otimes |B;j'_l\rangle$, and the second equality 
holds for an overall pure state with only one term contributing to 
the sum over $l$, hence $l$-dependence is suppressed; 
note that each one of the weights $w$ has the 
meaning of a probability. The interpretation here is an obvious generalization of 
the one following Eq.\,(\ref{rhointerpr2}). 

Next, we distinguish pure and mixed entangled states of the QM composite 
system. -- The case of {\it pure entangled states} can be easily fit into the above 
considerations. To the bi-partite (finite-dimensional) Hilbert space, in which a 
pure state has a given tensor product structure, corresponds  
a factorized algebra of QM observables. It has been shown recently that this situation 
is unitarily equivalent to one where the Hilbert space is unstructured, provided 
the algebra of observables is suitably rearranged ({\it ``Tailored Observables Theorem''}) 
\cite{HarshmanRanade11}. Thus, the tensor product structure can be effectively 
eliminated and the entanglement properties shifted to the observables, and vice versa.  
With an unstructured Hilbert space, we are back to the situation discussed in 
the context of Eqs.\,(\ref{rhointerpr1})--(\ref{rhointerpr2}). In this way, 
in our formulation, pure entangled states from the QM sector can naturally occur in a 
QM-CL hybrid. 
 
Mixed states present a more subtle situation. This is immediately obvious if 
one looks at the limiting case of a {\it totally mixed state}, described by 
$\hat\rho =N^{-1}\sum_{i=1}^N |\Phi_i\rangle\langle\Phi_i|\equiv N^{-1}\mathbf{1}_N$, where $\{\Phi_i\}$ stands for  
any complete orthonormal basis of the $N$-dimensional Hilbert space of 
the composite. By definition, cf. Eq.\,(\ref{separable}), this state is separable for all 
(factorized) representations, and thus can be handled as before. 
By continuity, this holds for states in a sufficiently small 
neighbourhood of the totally mixed state. However, it turns out that the possibility (or not) 
to switch between entangled and separable forms of the state, by tailoring the algebra 
of observables, which we needed for our above arguments, still cannot be ascertained 
for mixed states in all generality. Criteria to assess this situation are presently 
under investigation, see Ref.\,\cite{ThirringEtAl11}, with references to earlier 
work therein. In conclusion, for {\it mixed entangled states} an intuitive interpretation of the 
hybrid density, along the lines sketched above, is still missing. Yet we are confident 
that this will work out eventually. 

\subsection{Evolution of hybrid densities}
Turning to dynamics, we proposed a straightforward 
generalization of Hamiltonian mechanics (see also Section~4.), 
which we employ here in the form of a {\it Liouville equation}, 
in order to describe the evolution of hybrid ensembles \cite{me11}. 
This is based on Liouville's theorem and the generalized Poisson bracket defined in 
Eqs.\,(\ref{GenPoissonBracket})--(\ref{GenPoissonBracketdef}) and assumes the 
compact form: 
\begin{equation}\label{rhoevol} 
-\partial_t\rho = \{\rho ,{\cal H}_\Sigma\}_\times  
\;\;, \end{equation}
with ${\cal H}_\Sigma\equiv{\cal H}_\Sigma (x_k,p_k;X_i,P_i)$ and:  
\begin{equation}\label{HtotalInt} 
{\cal H}_\Sigma:={\cal H}_{\mbox{\scriptsize CL}}(x_k,p_k)
+{\cal H}_{\mbox{\scriptsize QM}}(X_i,P_i) 
+{\cal I}(x_k,p_k;X_i,P_i)   
\;\;, \end{equation} 
which defines the relevant {\it Hamiltonian function}, including a hybrid interaction. 
We require ${\cal H}_\Sigma$ to be an {\it observable}, last not least to have 
a meaningful notion of conserved energy. 

An important advantage of Hamiltonian flow and a general property of the Liouville 
equation in this context is \cite{Arnold}:  
\begin{itemize} 
\item The normalization and positivity of the probability 
density $\rho$ are conserved in the presence of 
a genuine hybrid interaction. Therefore, the interpretation of $\rho$ as a 
probability density remains valid. 
\end{itemize}

However, the simple relation between $\rho$ as a function of the QM ``phase space'' 
variables $X_i,P_i$ and an expectation of a self-adjoint density operator $\hat\rho$, 
as employed so far, does not continue to hold generally under hybrid evolution. 
As pointed out in Section~5.4 of Ref.\,\cite{me11}, the oscillator 
expansion of observables, such as in the second of Eqs.\,(\ref{rhodens}), 
has to be replaced by a more general form, appropriate for what we named  
{\it almost-classical observables}, cf. below.   

This comes about, since the ``classical part'' of the bracket, 
$\{ A,B\}_{\mbox{\scriptsize CL}}$, can generate terms which do {\it not} qualify as 
observable with respect to the QM sector; here we assume that $A$ and 
$B$ are both hybrid observables, as defined before. 
Such terms are of the form: 
\begin{eqnarray} 
&\;&\frac{1}{4}\sum_{i,i',j,j'}\{A_{ij},B_{i'j'}\}_{\mbox{\scriptsize CL}} 
(X_i-iP_i)(X_j+iP_j)(X_{i'}-iP_{i'})(X_{j'}+iP_{j'}) 
\nonumber \\ [1ex] \label{nonlinear} 
&\;&=\sum_{i,i',j,j'}\langle\Psi |\Phi_i\rangle\langle\Psi |\Phi_{i'}\rangle 
\{A_{ij},B_{i'j'}\}_{\mbox{\scriptsize CL}} 
\langle\Phi_j|\Psi\rangle\langle\Phi_{j'}|\Psi\rangle
\;\;, \end{eqnarray} 
where we used the oscillator expansion, Eq.\,({\ref{oscillexp}), and: 
\begin{equation}\label{nonlinear1}
\{A_{ij},B_{i'j'}\}_{\mbox{\scriptsize CL}} 
=\sum_k
\Big (\frac{\partial A_{ij}}{\partial_{x_k}}\frac{\partial B_{i'j'}}{\partial_{p_k}}
-\frac{\partial A_{ij}}{\partial_{p_k}}\frac{\partial B_{i'j'}}{\partial_{x_k}}\Big )
\;\;, \end{equation} 
since, for example, 
$A\equiv A(x_k,p_k;X_i,P_i)=\sum_{i,j}A_{ij}(x_k,p_k)(X_i-iP_i)(X_j+iP_j)$, cf. 
Eq.\,(\ref{Gexp}). 

In this way, evolution of hybrid observables, of the density $\rho$ 
in particular, can induce a {\it structural change}: while continuing to be CL 
observables, they do not remain QM observables (quadratic forms in $X_i$'s and $P_i$'s). 
They fall outside of the product algebra generated by the 
observables to which we confined ourselves in Section~2. Therefore, 
if we wish to maintain the formal consistency of our scheme, we have to assume: 
\begin{itemize} 
\item The algebra of hybrid 
observables is closed under the QM-CL Poisson bracket operation, implemented 
by $\{\;,\;\}_\times$ . 
\end{itemize} 
Thus, the product algebra of CL and QM observables is 
replaced by its larger closure.  
This amounts to a {\it physical hypothesis}, as we shall discuss in the remainder 
of this section. 

Refering to the phase space coordinates $(X_i,P_i)$, 
we define an {\it almost-classical observable} as a real-valued regular function 
of pairs of factors like $(X_i-iP_i)(X_j+iP_j)$, such as in the 
left-hand side of Eq.\,(\ref{nonlinear}), subject to the constraint:   
${\cal C}(X_i,P_i)=\frac{1}{2}\sum_i(X_i^{\;2}+P_i^{\;2})\stackrel{!}{=}1$. 

This normalization constraint, cf. Eq.\,(\ref{oscillnormalization}), 
is preserved under the evolution, since $\{{\cal C},{\cal H}_\Sigma\}_\times =0$, 
in the presence of QM-CL hybrid interaction.  
Furthermore, consistently with the closure of the enlarged algebra of observables, 
we find:  
\begin{equation}\label{constraintG} 
\{{\cal C}(X_i,P_i),{\cal G}(x_k,p_k;X_i,P_i)\}_\times 
=\{{\cal C}(X_i,P_i),{\cal G}(x_k,p_k;X_i,P_i)\}_{\mbox{\scriptsize QM}} 
=0 
\;\;, \end{equation} 
where ${\cal G}(x_k,p_k;X_i,P_i)$ stands for any almost-classical observable, 
including QM observables as a special case, of course. This follows, since 
the explicit form of the generalized observables, as on the left-hand side of 
Eq.\,(\ref{nonlinear}), or generalizations including additional pairs of 
factors like $(X_i-iP_i)(X_j+iP_j)$, leads to a sum of terms, each one vanishing 
as if it stemmed from a QM observable, which commute with the constraint represented by 
${\cal C}$ under the Poisson bracket.   

According to this definition, members of the complete algebra of hybrid observables, 
generally, are {\it classical} with respect to 
coordinates $(x_k,p_k)$ and {\it almost-classical} with respect to coordinates $(X_i,P_i)$, 
which can be restated as:  
\begin{itemize} 
\item QM observables (quadratic forms in phase space coordinates) 
form a subset of almost-classical observables which, in turn, form a subset of 
classical observables 
(real-valued regular functions of phase space coordinates), cf. Section~2.  
\end{itemize} 

Physical consequences of this enlarged 
``classical$\;\times\;$almost-classical algebra'' for interacting QM-CL hybrids 
are illustrated by the following {\it Gedankenexperiment}. 

Consider a quantum together with a classical object subject to a transient hybrid interaction. 
As long as the hybrid interaction is ineffective, both objects 
evolve independently according to Schr\"odinger's and Hamilton's equations, respectively. 
However, once they form an interacting hybrid, the corresponding 
phase space density changes from a factorized form, in absence of any initial 
correlation, to become an 
almost-classical/classical hybrid observable. Most likely, the density maintains such a mixed 
character, even when the hybrid interaction ceases. 

This outcome contradicts naive expectation that quantum and classical objects 
evolve separately in quantum and classical ways, when they no longer 
interact. -- Two possibilities come to mind. Either persistence of the 
almost-classical/classical character is a {\it physical 
effect} accompanying QM-CL hybrids, if they exist. Or our description  
needs to be augmented with a {\it reduction mechanism} by which evolving observables return to  
standard QM or CL form, following a hybrid interaction.  
Both possibilities seem quite interesting in their own right. 

Presently, we take the enlargement of the QM algebra 
of observables, induced by a hybrid interaction, as {\bf ``first hint''} that features of  
QM-CL hybrids might be relevant for an understanding of {\it how QM emerges}. 
One would like to understand how a large  
algebra of classical observables (functions on phase space) is reduced, 
possibly via almost-classical observables at an intermediary stage, to a smaller QM algebra 
(linear operators on Hilbert space) for an object that becomes ``quantized''.  

\section{Does hybrid dynamics allow for ``Free Will''?} 
In order to address this question, we recall the generalized Hamiltonian equations 
of motion derived earlier \cite{me11}. They reflect the dynamical structure 
underlying hybrid evolution, according to our proposal, and correspond in detail to 
the Liouville equation of the previous section. 

We consider hybrid systems described by a generic classical {\it Hamiltonian function}  
and a quantum mechanical {\it Hamiltonian operator}, respectively: 
\begin{eqnarray}\label{Hcl}  
{\cal H}_{\mbox{\scriptsize CL}}&:=&\sum_k\frac{p_k^{\; 2}}{2}+v(x_l)  
\;\;, 
\\ [1ex] \label{Hqm} 
\hat H_{\mbox{\scriptsize QM}}&:=&\frac{\hat P^2}{2}+V(\hat X)
\;\;, \end{eqnarray} 
where $v(x_l)\equiv v(x_1,\dots,x_n)$ and $V$ denote relevant potentials 
and all masses are set equal to one, for simplicity. Furthermore, 
there is a {\it self-adjoint hybrid 
interaction operator} $\hat I(x_k,p_k;\hat X,\hat P)$, invoking symmetrical (Weyl) 
ordering of the noncommuting operators $\hat X$ and $\hat P$. 
By Eq.\,(\ref{goperator}), this gives rise 
to the following {\it Hamiltonian function} ${\cal H}_\Sigma$: 
\begin{eqnarray}
{\cal H}_\Sigma =\sum_k\frac{p_k^{\; 2}}{2}+v(x_l)
+\langle\Psi |\Big (\frac{\hat P^2}{2}+V(\hat x)\Big )|\Psi\rangle 
+\langle\Psi |\hat I(x_k,p_k;\hat X,\hat P)|\Psi\rangle 
\nonumber \\ [1ex] \label{HSigmaGen1} 
\;\;\;\;\;\;\; =:\; {\cal H}_{\mbox{\scriptsize CL}}(x_k,p_k)
+{\cal H}_{\mbox{\scriptsize QM}}(X_i,P_i) 
+{\cal I}(x_k,p_k;X_i,P_i)   
\;\;, \end{eqnarray} 
when evaluated in a pure state $|\Psi\rangle$, invoking the oscillator representation  
of Eq.\,(\ref{oscillexp}). 
With these definitions in place, one finds the equations of motion 
by the rules of Hamiltonian dynamics. 

The equations of motion for the CL {\it observables} $x_k,p_k$ are: 
\begin{eqnarray}\label{xdot} 
\dot x_k&=&\{ x_k,{\cal H}_\Sigma\}_\times 
=p_k+\partial_{p_k}{\cal I}(x_k,p_k;X_i,P_i)
\;\;, \\ [1ex] \label{pdot} 
\dot p_k&=&\{ p_k,{\cal H}_\Sigma\}_\times 
=-\partial_{x_k}v(x_l)-\partial_{x_k}{\cal I}(x_k,p_k;X_i,P_i) 
\;\;. \end{eqnarray} 
Similarly, we obtain for the QM {\it variables} $X_i,P_i$, which are {\it not observables}: 
\begin{eqnarray}\label{Xidot} 
\dot X_i&=&\{ X_i,{\cal H}_\Sigma\}_\times 
=\partial_{P_i}{\cal H}_{\mbox{\scriptsize QM}}(X_j,P_j)+\partial_{P_i}{\cal I}(x_k,p_k;X_j,P_j)
\\ [1ex] \label{Xidot1} 
&=&E_iP_i+\partial_{P_i}{\cal I}(x_k,p_k;X_j,P_j) 
\;\;, \\ [1ex] \label{Pidot} 
\dot P_i&=&\{ P_i,{\cal H}_\Sigma\}_\times 
=-\partial_{X_i}{\cal H}_{\mbox{\scriptsize QM}}(X_j,P_j)-\partial_{X_i}{\cal I}(x_k,p_k;X_j,P_j) 
\\ [1ex] \label{Pidot1} 
&=&-E_iX_i-\partial_{X_i}{\cal I}(x_k,p_k;X_j,P_j)
\;\;, \end{eqnarray} 
where Eqs.\,(\ref{Xidot1}) and (\ref{Pidot1}) follow, if the oscillator expansion 
is performed with respect to the stationary states of $\hat H_{\mbox{\scriptsize QM}}$, 
cf. Eqs.\,(\ref{HamiltonianQM1})--(\ref{HamiltonianQM2}). 
 
Notably, the Eqs.\,(\ref{xdot}), (\ref{pdot}) together with Eqs.\,(\ref{Xidot}), (\ref{Pidot}),  
or together with Eqs.\,(\ref{Xidot1}), (\ref{Pidot1}), form a {\it closed set} of $2(n+N)$ 
deterministic equations, where $n$ denotes the number of CL degrees of freedom and $N$ the 
dimension of the QM Hilbert space (assumed denumerable, if not finite).  
Earlier we contrasted this exact result with {\it generalized 
Ehrenfest equations} obtained for the hybrid model \cite{me11}, which we will 
employ in Section~5. 

Following Di\'osi \cite{Diosi11}, someone might ask ``Do I have the freedom 
({\it ``Free Will''})
to measure one of the CL observables, say $z$, and, conditioned on its value, 
to perturb the subsequent hybrid evolution in such a way that the following 
reasonable properties are guaranteed?'': \begin{itemize}  
\item i) the variable $z$ is a smooth real function of time;  
\item ii) the CL and QM variables, as well as the density $\rho$ (with its statistical interpretation), coexist and depend on each other;  
\item iii) the perturbation at time $t_0$ affects the QM-CL hybrid in a causal way, i.e., 
it has no effect at earlier times $t<t_0$. 
\end{itemize}  
The classical variable is called {\it tangible}, in this case.  

In the present hybrid theory the CL observables pertain to a classical object with 
its own dynamics, in the absence of interaction with a quantum object; 
they are {\it not} thought to 
derive somehow from measurements performed on this or an auxiliary 
quantum object, unlike in all models discussed by Di\'osi. So measuring $z$ could be 
the result of a coupling to another CL object, which serves as {\it apparatus}. 
 
In this case, it seems clear that property i) holds, provided the perturbation applied to 
the hybrid, in the form of some $\delta_z(t)$, added to $z$ for $t\ge t_0$, is sufficiently 
smooth; the effect on the equations of motion, Eqs.\,(\ref{xdot})--(\ref{Pidot1}), is 
simply to change initial conditions of the subsequent evolution. Similarly, 
properties ii) and iii) hold as a property of the Hamiltonian dynamics considered; 
consequently, this applies for the density $\rho$, evolving according to the equivalent 
Liouville equation, Eq.\,(\ref{rhoevol}), as we discussed in Subsection~3.2. 

We conclude that there is ``Free Will'' and classical observables are {\it tangible}  
in our theory.   

\section{Does locality remain intact in hybrids?}  
Let us consider the Ehrenfest relations for hybrids \cite{me11}. 
Here we look at the case of bilinearly coupled oscillators, 
which yields a very simple version of these equations and still illustrates  
the issue of locality. -- The following discussion applies as well to 
the general equations of motion of Section~4, as we shall see shortly. 

Earlier we obtained the closed set of equations determining the 
CL observables $x_k,p_k$ 
and QM {\it coordinate and momentum observables} $X,P$, defined by:  
\begin{equation}\label{QMobservables} 
X(X_i,P_i):=\langle\Psi |\hat X|\Psi\rangle \;\;,\;\;\;
P(X_i,P_i):=\langle\Psi |\hat P|\Psi\rangle 
\;\;, \end{equation}
following our construction, cf. Section~2., and   
assuming here a {\it set of} CL oscillators coupled bilinearly 
to {\it one} QM oscillator, as defined by: 
\begin{eqnarray}\label{Hcl1}  
{\cal H}_{\mbox{\scriptsize CL}}&:=&\sum_k
\Big ({\textstyle \frac{1}{2m_k}p_k^{\; 2}}
+{\textstyle \frac{m_k\omega_k^{\; 2}}{2}}x_k^{\; 2}\Big ) 
\;\;, 
\\ [1ex] \label{Hqm1} 
\hat H_{\mbox{\scriptsize QM}}&:=&{\textstyle \frac{1}{2M}}\hat P^2
+{\textstyle \frac{M\Omega^2}{2}}\hat X^2 
\;\;, 
\\ [1ex] \label{bilinear} 
\;\;\;\;\;\;\hat I&:=&\hat X\sum_k\lambda_kx_k
\;\;. \end{eqnarray} 
The constants $m_k,M$, $\omega_k,\Omega$, and $\lambda_k$ denote 
masses, frequencies, and couplings, respectively. -- 
For this hybrid system, the general equations of motion and Ehrenfest 
relations indeed reduce to a simple {\it closed} set of equations:  
\begin{eqnarray}\label{xdot1} 
\dot x_k&=&{\textstyle \frac{1}{m_k}}p_k
\;\;, \\ [1ex] \label{pdot1} 
\dot p_k&=&-{\textstyle m_k\omega_k^{\; 2}}x_k 
-\lambda_k X 
\;\;, \\ [1ex] \label{Xdot1} 
\dot X&=&{\textstyle \frac{1}{M}}P
\;\;, \\ [1ex] \label{Pdot1}
\dot P&=&-{\textstyle M\Omega^2}X
-\sum_k\lambda_kx_k 
\;\;. \end{eqnarray} 
We observe that the {\it backreaction} of QM on CL subsystem appears, {\it as if} 
the CL subsystem was coupled to another CL oscillator. For a more general discussion 
of backreaction of QM on CL subsystems, see Ref.\,\cite{me11}, in particular, concerning 
the role of quantum fluctuations. 

However, as defined above, the QM observables are expectations and we have 
explicitly, with $\hat X|q\rangle =q|q\rangle$, for example: 
\begin{eqnarray} 
X(X_i,P_i)&=&\int\mbox{d}q'\mbox{d}q\;\langle\Psi |q'\rangle\langle q'|\hat X|
q\rangle\langle q|\Psi \rangle 
=\int\mbox{d}q\;q|\Psi (q)|^2 
\nonumber \\ [1ex] \label{Xexpect}
&=&\frac{1}{2}\sum_{i,j}(X_i-iP_i)(X_j+iP_j)\int\mbox{d}q\;
\langle\Phi_i|q\rangle q\langle q|\Phi_j\rangle  
\;\;, \end{eqnarray} 
or, more generally, for hybrid interactions that 
include typical position measurement interactions: 
\begin{equation}\label{Iexpect}
{\cal I}(x_k,p_k;X_i,P_i)=\frac{1}{2}\sum_{i,j}(X_i-iP_i)(X_j+iP_j)\int\mbox{d}q\;
\Phi^\ast _i(q)\Phi_j(q){\cal I}(x_k,p_k;q)  
\;\;, \end{equation} 
with $\{ |\Phi_j\rangle\}$ representing any complete orthonormal set of states. 
According to quantum theory, if the QM subsystem of the hybrid is evolving 
in spacetime, the integration over 
the variable $q$ is understood as an {\it integration over all space}. Thus, the 
coupling in Eqs.\,(\ref{HSigmaGen1}) or (\ref{bilinear}) is generally 
{\it highly nonlocal} indeed. 

This nonlocality has been held as a deficiency against mean-field semiclassical 
theories and the present theory, as discussed so far, shares this problem. 

However, several remarks are in order here. -- First of all, we may introduce a
localizing factor and redefine the hybrid interaction: 
\begin{eqnarray}\label{Ilocal} 
{\cal I}(x_k,p_k;X_i,P_i)&:=&\langle\Psi |\hat {\cal I}(x_k,p_k;\hat X)
\delta (\hat X-x_k\hat \mathbf{1})|\Psi\rangle 
\\ [1ex] \label{Ilocal1}  
&=&\frac{1}{2}\sum_{i,j}(X_i-iP_i)(X_j+iP_j)
\Phi^\ast _i(x_k)\Phi_j(x_k){\cal I}(x_k,p_k;x_k) 
\;\;. \end{eqnarray} 
For the bilinear interaction of Eq.\,(\ref{bilinear}), for example,  
we obtain here by localization: 
\begin{eqnarray} 
{\cal I}(x_k;X_i,P_i)&=&\sum_{k}\lambda_kx_k\cdot x_k|\Psi (x_k)|^2
\nonumber \\ [1ex] \label{bilinear1} 
&=&\frac{1}{2}\sum_{i,j,k}
\lambda_kx_k^{\;2}\Phi^\ast _i(x_k)\Phi_j(x_k)(X_i-iP_i)(X_j+iP_j)
\;\;, \end{eqnarray} 
where the couplings $\lambda_k$ include a suitable dimensional factor, as compared to before. 

This definition of the hybrid interaction is consistent 
with our formal framework and solves the locality issue. It describes the 
interaction in terms of the CL coordinates, however, weighted by appropriate 
wave function factors. The latter reflect that the QM subsystem is generally 
not localized. Of course, more general forms of the localizing factor are possible and  
can be extended to field theories as well. 
  
By localizing the hybrid interaction in this way, we {\it assume ad hoc} that the QM subsystem 
is probed by the CL subsystem at its location. 

Due to the nonlinear features of the 
localization, the generalized Ehrenfest relations obtain additional terms, as compared 
to our earlier results \cite{me11}, which we will not explore here. 

Instead, we would like to draw attention to an aspect of {\it locality of hybrid dynamics}, 
in particular of the general equations of motion, Eqs.\,(\ref{xdot})--(\ref{Pidot1}).  
This will turn out to be independent of whether localization, as discussed above, is 
present or not.  

Suppose a physicist lacking any knowledge of quantum mechanics were presented with these 
equations (plus the normalization constraint, Eq.\,(\ref{oscillnormalization})). -- {\it We 
know} that these equations present independent CL 
and QM sectors, in the absence of a hybrid interaction. -- However, he/she would naturally 
interpret them to describe the dynamics of a 
composite CL object, with part of its phase space compactified (due to the constraint). 
Looking at it this way, {\it he/she finds a perfectly local dynamics}.  
In fact, {\it our} knowledge of nonlocal features can be traced to the definition of   
the canonical coordinates and momenta $X_i,P_i$, introduced by the oscillator representation, 
Eq.\,(\ref{oscillexp}), since: $X_i/\sqrt 2=\mbox{Re}\int\mbox{d}q\;\Phi^\ast _i(q)\Psi(q)$ 
and $P_i/\sqrt 2=\mbox{Im}\int\mbox{d}q\;\Phi^\ast _i(q)\Psi(q)$. Therefore, 
spatially nonlocal (and probabilistic) features enter only by reference to the 
QM wave function. 

We tentatively conclude that the spatial interpretation of QM phenomena is, in some sense 
to be better understood, of a secondary character and take this as {\bf ``second hint''} 
that features of QM-CL hybrids might be relevant for an understanding of {\it how QM emerges}.

\section{Are there hints for the emergence of quantum mechanics?}    
Our presentation of QM-CL hybrid dynamics has been based on suitably re-presenting 
the QM sector in the generalized analytical mechanics framework incorporating quantum 
mechanics, which was outlined by Heslot \cite{Heslot85}. This has allowed us, 
in particular, to generate the hybrid evolution by a generalized Poisson bracket 
and to have a unified QM-CL description of states and observables. 

As we discussed in Sections~3. and 5., we have found at least two hints that 
our present description may have already something to say about dynamics beyond quantum,  
classical, and QM-CL hybrid mechanics, indicating structures that 
lead us beneath quantum mechanics.   

A caveat is in order here. Dynamics from 
which quantum mechanics could emerge by some coarse-graining process for large scales 
will not necessarily have the Hamiltonian structure that we invoked. It could be of a much 
more general, less structured kind, such as based on discrete deterministic automata 
\cite{tHooft10}.  

However, the present work could help to identify deviations from quantum mechanics, 
which seemingly works perfectly on the scales accessible to present-day experiments. 
For this purpose, the Hamiltonian dynamics here can serve as a model, in order to 
parametrize such deviations.  

It has been repeatedly pointed out by V.I.\,Man'ko and collaborators that classical 
states may differ widely from what could be obtained as the ``$\hbar\rightarrow 0$'' 
limit of quantum mechanical ones. Furthermore, they show that all states can 
be classified by their ``tomograms'' as {\it either} CL {\it or} QM, CL {\it and} QM, 
and {\it neither} CL {\it nor} QM~\cite{Manko04,Manko11}. 
Yet, in order to explain these {\it ``Man'ko classes of states''}, 
some unknown dynamics seems missing.

In this context, we find the {\bf ``first hint''} of Section~3. very interesting, namely 
that a consistent hybrid description forces an enlarged algebra of observables upon us, 
as compared to the quantum mechanical one. 
More specifically, the QM observables obtain through a 
{\it shrinking of the algebra of observables}, 
via intermediary almost-classical observables, from the largest algebra of classical 
observables. This seems to complement and underline the findings by Man'ko in our approach, 
where all states are represented in a suitable phase space. We have also 
emphasized the {\it spherical compactification of QM state space}, which is absent 
classically. 

It would be most interesting 
to find underlying dynamical reasons for such {\it structural change} that occurs when 
a CL object interacts with a QM object. This might help to explain how quantum mechanics 
emerges in a deterministic world.      

Furthermore, a {\bf ``second hint''}, concerning the relevance of QM-CL hybrids for an understanding of how QM emerges, is coming from Section~5. There, we have seen that the 
hybrid dynamics might appear as a perfectly local deterministic scheme, 
if it were not for seemingly added-on interpretation of the QM sector with the help 
of a wave function defined on its to-be configuration space. Let us assume that spatial 
notions are somehow related to how gravity enters the picture. Then, leaving the latter 
aside at this time, 
one might devise a rather general dynamical framework, possibly incorporating 
dissipation in the form of an attractor mechanism, already alluded to in 
Refs.\,\cite{Vitiello01,tHooft06a,Elze08,Blasone09}, which drives classical through 
hybrid to quantum mechanics.  

\section{Concluding remarks} 
We have discussed four questions in these notes which can be addressed to any 
proposal of a quantum-classical hybrid theory, in addition to the consistency 
requirements listed in the introduction, which, in turn, have been widely studied 
by others. Presently, we considered these questions in the context 
of the hybrid theory of Ref.\,\cite{me11}, which meets all consistency requirements 
and can be seen as a generalization of proposals incorporating mean field theory.  

We emphasize that here the classical sector is not necessarily meant to present  
an approximation for some of the quantum mechanical degrees of freedom in a 
fully quantum mechanical multi-partite system. Rather, 
we have continued to study here, as well as earlier, 
whether such a hybrid model as ours can stand formally on 
its own and meet precisely all the consistency requirements posed.    

Thus, the questions raised at present are meant to further illuminate the contents and 
limitations of our or similar models. 

The {\it first question} essentially asks, whether the proposed formalism is general enough 
to incorporate all possible quantum states, including superpositions, entangled or 
separable pure and mixed states. Which we have answered positively, with the exception 
of mixed entangled states, where the necessary tailoring of observables has not yet  
been achieved in quantum mechanics proper in all generality, cf. 
Ref.\,\cite{ThirringEtAl11}. We have interpreted (Section~3.) the encountered necessity of 
an enlarged algebra of almost-classical observables as a {\bf ``first hint''} 
that hybrid dynamics might offer a glimpse beyond quantum mechanics.  

The {\it second question} concerns the control of a hybrid system by external classical 
means, obviously a question of great practical importance. Indeed, our proposal 
allows for this and for ``Free Will'', in particular, in the sense defined by 
Di\'osi \cite{Diosi11}. 

The {\it third question}, whether locality is not violated by the proposed hybrid dynamics, 
as in other proposals related to mean field theory, has been discussed in some detail. 
Only if we invoke an ad hoc localization of the relevant interactions, this seems 
to be guaranteed. On the other hand, leaving out the interpretation of the quantum 
mechanical sector related to a wave function on configuration space, we have argued 
that here we may have an interesting indication of dynamics 
-- a {\bf ``second hint''} -- that transcends the usual quantum mechanical framework.

In Section~5., we outlined what might be taken forward from quantum-classical 
hybrid dynamics tackling the emergence of quantum mechanics from a deterministic theory 
beneath, which comprises the {\it fourth question}. 
Based on the formal developments of Ref.\,\cite{me11} and further 
discussion here, we seem to get closer to be able to study these issues in some detailed models.    

\ack{It is a pleasure to thank L.~Di\'osi, P.~Hajicek, M.J.W. Hall, and M. Reginatto 
for asking related questions and G.~Gr\"ossing and co-organizers for inviting me to 
the {\it ``Heinz von Foerster Congress / Emergent Quantum Mechanics''}.}~\footnote{After 
completion of this work, we have noticed Ref.\,\cite{Salcedo12}, where issues of  
Section~3. are discussed in different ways. A stringent requirement of {\it statistical consistency} is introduced which is, however, based on the assumption that its validity 
extends from QM to quantum-classical hybrid dynamics.} 


\end{document}